\documentstyle{article}
\begin{document}
\fontsize{10 pt}{13 pt}
\selectfont
\begin{center}
{\bf ENERGY DISTRIBUTION OF A CHARGED BLACK HOLE WITH A MINIMALLY COUPLED SCALAR FIELD} \vspace{1 cm}

\fontsize{8pt}{10pt} \selectfont PAUL HALPERN \\
\vspace{2 pt}
{\it Department of Mathematics, Physics and Statistics,\\
University of the Sciences in Philadelphia, 600 S. 43rd St.,\\
Philadelphia, PA. 19104, USA\\ p.halper@usip.edu}
\end{center}
\begin{quote}
\fontsize{8pt}{10pt} \selectfont Using three different
energy-momentum complexes, the Einstein, Landau-Lifshitz, and
Papapetrou prescriptions, we calculate the energy of an electrically
charged black hole exact solution with a self-interacting,
minimally-coupled scalar field and the asymptotic region locally an
anti-deSitter spacetime.  Writing the metric in Kerr-Schild
Cartesian coordinates, we demonstrate that this metric belongs to
the Kerr-Schild class of solutions.  Applying each of the three
energy-momentum prescriptions and comparing the results, we find
consistency among these complexes, suggesting their utility as
localized measures of energy.


\end{quote}
\fontsize{10 pt}{13 pt} \selectfont \vspace{1 cm} \noindent
 {\bf 1. Introduction}

\vspace{6 pt} One of Einstein's challenges in formulating the
general theory of relativity was incorporating the energy and
momenta of gravitational fields along with those of matter and
non-gravitational fields into a local, well-defined conserved
quantity [1]. Einstein's efforts along these lines led him to
propose an energy-momentum complex that would include all fields and
obey a conservation law [2].  Non-tensorial in nature, the Einstein
complex is highly dependent on the coordinate system used. In the
case of a flat, empty space-time, for example, the Einstein complex
vanishes for quasi-Cartesian coordinates  but is infinite for
spherical coordinates.  

The Einstein energy-momentum complex is asymmetric
 in its indices, precluding the definition of a conserved angular
 momentum and leading Landau and Lifshitz [3],
 Papapetrou [4], Weinberg [5] and others
 to construct alternative formulations.  In general, these favor
 quasi-Cartesian coordinate systems.  Yet another prescription,
 developed by M{\o}ller [6], is uniquely designed to be independent
  of the choice of coordinate system.

The multiplicity of energy-momentum complexes and their particular
constraints once deterred theorists from their use.  However, in
recent years interest has been revived due to a number of promising
findings about consistent ways of defining energy localization.  In
1990 Bondi argued that energy could at least in principle be
localized [7].  The same year
 Virbhadra found that different energy-momentum complexes
when applied to the same metric furnish reasonable, consistent
results [8].

These results were extended in 1996 when
Aguirregabiria, Chamorro and Virbhadra established that all metrics
of the Kerr-Schild class
 (the Schwarzchild, Reissner-Nordstr{\"o}m and many other well-known
 solutions) have unique, well-defined energy and momentum distributions even
when different complexes are applied.  It remains an open question why certain 
classes of metrics yield identical energy-momentum distributions for a range of 
complexes, while others do not produce such consistency.  Nevertheless, as 
Aguirregabiria, Chamorro and Virbhadra wrote, 
``different pseudotensors giving the same results for local quantities... does not
seem to be accidental,'' [9] suggesting that certain symmetries present in particular metrics
lend themselves well to consistent definitions of local conservation laws.  Resolving this issue 
has been the subject of continuing investigation, including cosmological as well as
stationary solutions.

In 1994 Rosen applied the Einstein energy-momentum complex to a
closed Friedmann-Robertson-Walker (FRW) cosmology, found a zero
value for its energy integrated over the full range of spatial
coordinates, and speculated that the total energy of the universe is
identically zero [10].  Other researchers
 have extended these results to other complexes [11],  anisotropic
 cosmologies  [12-17],
 and teleparallel gravity (a modification of general relativity) [18-21].

Black holes and related objects have been a much studied application
of energy-momentum complexes.  Using various complexes, Vagenas has
examined the dyadosphere of Reissner-Nordstr{\"o}m black holes [22],
investigated stringy black
holes [23], considered the Ba–ados, Teitelboim, and Zanelli (BTZ) metric 
(a rotating 2+1 dimensional black hole solution) [24], and
evaluated the energy of radiating charged particles [25]. Xulu has
also calculated the energy of
 stringy black holes [26].  Radinschi and her
colleagues  [27, 28], as well as Gad [29], have also investigated a
variety of cases for 2+1 dimensional and stringy black holes.
Chamorro and Virbhardra [30], Vagenas [31] and Xulu [32] have each
calculated the energy distribution of charged black holes with a
dilaton field. Yet other researchers have looked at black plane
solutions [33] and a stationary beam of light [34, 35]

Much of this work has been stimulated by a 1999 paper by Virbhadra
suggesting that the energy-momentum complexes of Einstein, Landau
and Lifshitz, Papapetrou and Weinberg yield the same results for
metrics more general than the Kerr-Schild class if they are
expressed in terms of Kerr-Schild Cartesian coordinates [36].   Virbhadra 
emphasized that although the possibility of a local definition of energy and 
momentum is subject to debate and these complexes are not covariant, 
nevertheless they offer promise because of such consistency and also due to
the fact that they each satisfy local conservation laws in all coordinate systems.
More recently, Xulu has extended this work by applying the
Bergmann-Thomson energy-momentum complex to solutions more general
than the Kerr-Schild class [37].

Along these lines it is interesting to apply several different
energy-momentum complexes to a newly found solution describing
a charged black hole with scalar ``hair.''

\fontsize{10 pt}{13 pt} \selectfont \vspace{1 cm} \noindent
 {\bf 2. Charged black holes with scalar hair}
\vspace{6 pt}

In the early 1970s, Wheeler and others, in considering stable black
hole solutions of the Einstein-Maxwell equations, formulated the
well-known ``no-hair theorem'' (from Wheeler's maxim: ``Black holes
have no hair'') limiting the observable properties of a black hole
to three parameters:  mass, electric charge and angular momentum
[38-40]. The black hole's event horizon cloaks any additional
information that could betray its origin.

In recent years, however, certain exceptions to the no hair theorem
have been revealed.  Numerical solutions that circumvent its
restrictions were found by Torii, Maeda and Narita, Winstanley and
others [41-44].  Exact hairy solutions with a scalar field were
discovered by Mart{\'i}nez, Troncoso and Zanelli [45] and later by
Zloshchastiev [46]. Koutsoumbas, Musiri, Papantonopoulos and Siopsis
explored the perturbative behavior of black holes with scalar hair
and identified a second-order phase transition [47].  Radu and
Winstanley argued for the existence of solutions with a conformally
coupled scalar field in four or more dimensions if a negative
cosmological constant is present [48].

In 2006, Mart{\'i}nez and Troncoso identified a new exact solution
describing a black hole in 3+1 dimensions with electric charge and a
self-interacting scalar field minimally coupled to electromagnetism
and gravitation [49].  The scalar field generates an effective
negative cosmological constant, rendering the black hole's event
horizon a surface of constant negative curvature, and making its
asymptotic behavior locally that of an anti-deSitter (AdS)
 spacetime.  An anti-deSitter spacetime is a maximally symmetric solution with a negative cosmological
 constant and possessing constant negative curvature.   A number of researchers have studied the thermodynamic properties of black holes
 with asymptotic AdS regions [50-54].
 In line with these earlier findings, Mart{\'i}nez and Troncoso have established that their solution
 has a well-defined fixed temperature, independent of horizon size, and an
 entropy that obeys the usual area law.  Their solution is massless, regardless of the value of the
 charge.  Although the influence of the scalar field represents a circumvention of the no-hair theorem,
 Mart{\'i}nez and Troncoso found that their solution has a finite probability of decaying into a black
 hole without a scalar field and hence without hair.

 Mart{\'i}nez and Troncoso's solution obeys the standard Einstein-Maxwell field equations modified by the addition of a minimally coupled, self-interacting scalar field, namely:

 \begin{eqnarray}
 G_{\mu \nu} &=& 8 \pi G ( {T_{\mu \nu}}^{(EM)} + {T_{\mu \nu}}^{(\phi)} )\\
{ \partial}_\nu (\sqrt{-g} F^{\mu \nu}) &=& 0\\
 \Box \phi &=& \frac{d V}{d \phi}
 \end{eqnarray}

where the electromagnetic and scalar parts of the stress-energy tensor are:

\begin{eqnarray}
{T_{\mu \nu}}^{(EM)}&=& -\frac{1}{4 \pi} (F^{\mu}_{\alpha}F^{\alpha \mu} +
\frac{1}{4} g^{\mu \nu} F^{\alpha \beta} F_{\alpha \beta})\\
{T_{\mu \nu}}^{(\phi)} &=& {\partial}_{\mu} \phi {\partial}_{\nu}
\phi - \frac{1}{2} g_{\mu \nu} g^{\alpha \beta} {\partial}_{\alpha}
\phi {\partial}_{\beta} \phi - g_{\mu \nu} V(\phi)
\end{eqnarray}

and the potential for the self-interacting scalar field has the
form:

\begin{equation}
V(\phi) = - \frac{3} {8 \pi G l^2} {\cosh}^4 (\sqrt{\frac{4 \pi G}{3} }\phi)
\end{equation}

For $\phi = 0$ this potential reaches a global maximum. The non-zero
expectation value yields an effective negative cosmological constant
that can be expressed in terms of the AdS radius as:

\begin{equation}
\Lambda = - \frac{3}{l^2}
\end{equation}

The field equations are satisfied by the metric:

\begin{eqnarray}
ds^2 &=& {(1 + \frac{G q^2}{r^2})}^{-1} \hspace{0.1 cm}
(\frac{r^2}{l^2} -1 + \frac{G q^2}{l^2}) \hspace{0.1cm}{dt}^2 \nonumber \\
&-& {(1 + \frac{G q^2}{r^2})}^{-2} \hspace{0.1 cm} (\frac{r^2}{l^2}
-1 + \frac{G q^2}{l^2})^{-1} \hspace{0.1 cm} {d r}^2 - r^2
{d\sigma}^2
\end{eqnarray}

with $q$ representing the electric charge and the
electromagnetic potential and scalar field set to be:

\begin{eqnarray}
A &=& - \frac{q}{\sqrt{r^2 + G q^2}} dt \\
\phi &=& \sqrt{\frac{3}{4 \pi G}} \hspace{0.1 cm} {\rm arctanh} \sqrt{\frac{Gq^2}{r^2 + G q^2}}
\end{eqnarray}

\vspace{1 cm} \noindent {\bf 3. Evaluating the energy using Einstein's energy-momentum complex}
\vspace{6 pt}

It is interesting to use Einstein's prescription to evaluate the
energy distribution of Mart{\'i}nez and Troncoso's metric.  Along
the lines of Virbhadra's findings in [40], we express the metric in
Kerr-Schild Cartesian coordinates, choosing natural units such that
$G = 1$.  We define a new time coordinate $T$ such that:

\begin{equation}
dT = {(1 + \frac{q^2}{r^2})}^{-\frac{3}{2}} \hspace{0.1 cm} dt +
[{(1 + \frac{q^2}{r^2})}^{-2}\hspace{0.1 cm} {(\frac{r^2}{l^2} -1 +
\frac{q^2}{l^2})}^{-1} - 1 ] \hspace{0.1 cm} dr
\end{equation}

The metric (8) can now be written in Kerr-Schild-Cartesian form as:

\begin{equation}
{ds}^2 = {dT}^2 + [{(1 + \frac{q^2}{r^2})}^{2} \hspace{0.1 cm}
(\frac{r^2}{l^2} -1 + \frac{q^2}{l^2}) - 1 ]{[dT + dr]}^2 - {dx}^2 -
{dy}^2 - {dz}^2
\end{equation}

with the three Cartesian spatial coordinates expressed in terms of
the spherical coordinates in the usual way:

\begin{eqnarray}
x &=& r \hspace{0.1 cm} \sin \theta \hspace{0.1 cm} \cos \phi \\
y &=& r \hspace{0.1 cm} \sin \theta \hspace{0.1 cm} \sin \phi \\
z &=& r \hspace{0.1 cm} \cos \theta
\end{eqnarray}

Einstein's prescription defines the local energy-momentum density
as:

\begin{equation}
{{\theta}_i}^k = \frac{1}{16 \pi} {{H_i}^{kl}}_{,l}
\end{equation}

where the superpotentials ${H_i}^{kl}$ are given by:

\begin{equation}
{H_i}^{kl} = \frac{g_{in}}{\sqrt{-g}}{[-g (g^{kn} g^{lm} - g^{ln} g^{km})]}_{,m}
\end{equation}

This complex has the antisymmetric property that:
\begin{equation}
{H_i}^{kl} = -{H_i}^{lk}
\end{equation}

We substitute the metric components (12) into (17) and find that the
relevant values of the superpotentials are:

\begin{eqnarray}
{H_0}^{01} &=& \frac{q^4 + 2 q^2 r^2 + 2 r^4}{r^6}- \frac{2 x(q^2+r^2)^3}{l^2 r^6} \\
{H_0}^{02}&=& \frac{q^4 + 2 q^2 r^2 + 2 r^4}{r^6}- \frac{2 y(q^2+r^2)^3}{l^2 r^6} \\
{H_0}^{03} &=& \frac{q^4 + 2 q^2 r^2 + 2 r^4}{r^6}- \frac{2
z(q^2+r^2)^3}{l^2 r^6}
\end{eqnarray}

The energy-momentum components can be found by integrating the
energy-momentum density over the volume under consideration:
\begin{equation}
P_i = \int \int \int {\theta_i}^0 dx^1 dx^2 dx^3
\end{equation}

Through Gauss's theorem we can express this as a surface integral:

\begin{equation}
P_i = \frac{1}{16 \pi} \int \int {H_i}^{0 \alpha} {\mu}_{\alpha} dS
\end{equation}

where ${\mu}_{\alpha}$ is the outward unit vector normal to the
spherical surface element:
\begin{equation}
dS = r^2 \hspace{0.1 cm}\rm{sin}\theta \hspace{0.1 cm} d\theta
\hspace{0.1 cm} d\phi
\end{equation}

Evaluating this integral over the full coordinate range, we find the
total energy of Mart{\'i}nez and Troncoso's solution within a sphere of radius $r$  to be:

\begin{equation}
E= P_0 = \frac{q^4}{2 r^3} + \frac{q^2}{r} + \frac{r}{2} -
\frac{(q^2+r^2)^3}{2 l^2 r^3}
\end{equation}

\vspace{1 cm} \noindent
 {\bf 4. Determining the energy by use of the Landau-Lifshitz pseudotensor}

\vspace{0.5 cm}

We now turn to a second procedure for determining the local energy, the method of Landau and Lifshitz.
One critique of Einstein's prescription is that it is not symmetric in its indices, thus precluding the definition of an angular momentum conservation law.  Landau and Lifshitz strived to ameliorate this situation by finding a local measure of energy and momentum that is symmetric in its indices.  The measure they found, although a pseudotensor, has the useful property of vanishing locally in inertial frames when expressed in quasi-Cartesian coordinates.  When added to the stress-energy tensor it acts as a conserved current.  Hence conservation laws for energy-momentum and angular momentum are each well-defined.

Based on these constraints, Landau and Lifshitz determined the energy-momentum pseudotensor to be:

\begin{equation}
L^{ij} = \frac{1}{16 \pi} S_{,kl}^{ijkl}
\end{equation}

where the $L^{i0}$ represent the energy and momentum densities and:

\begin{equation}
S^{ijkl} = -g \hspace{0.1 cm} [g^{ij} g^{kl} - g^{ik} g^{jl}]
\end{equation}

In similar manner to we can use Gauss's theorem to express the total energy as a surface integral:

\begin{equation}
E = \frac{1}{16 \pi} \int \int  h^{00 \alpha}  {\mu}_{\alpha} dS
\end{equation}

where the superpotentials $h^{00 \alpha}$ are defined as:

\begin{equation}
h^{00 \alpha} = S^{00k \alpha}_{,k}
\end{equation}

Substituting the metric components (12) into (27) and using (29) we
find the relevant values of the
 superpotentials to be:

\begin{eqnarray}
h^{001} &=& \frac{q^4 + 2 q^2 r^2 + 2 r^4}{r^6}- \frac{2 x(q^2+r^2)^3}{l^2 r^6} \\
h^{002}&=& \frac{q^4 + 2 q^2 r^2 + 2 r^4}{r^6}- \frac{2 y(q^2+r^2)^3}{l^2 r^6} \\
h^{003} &=& \frac{q^4 + 2 q^2 r^2 + 2 r^4}{r^6}- \frac{2 z(q^2+r^2)^3}{l^2 r^6}
\end{eqnarray}

We integrate this expression over the full range of coordinates for a sphere of radius $r$ 
and determine the total energy using the Landau-Lifshitz complex to be:

 \begin{equation}
E= \frac{q^4}{2 r^3} + \frac{q^2}{r} + \frac{r}{2} -
\frac{(q^2+r^2)^3}{2 l^2 r^3}
\end{equation}

This is identical to the expression obtained using Einstein's prescription.

\vspace{1 cm} \noindent
 {\bf 5. Using Papapetrou's energy-momentum complex}

\vspace{0.5 cm} Another energy-momentum complex, developed by Papapetrou, shares with the Landau-Lifshitz complex the advantage of being symmetric in its indices and thereby allowing the conservation of angular momentum to be well defined.

Papapetrou defined his energy-momentum complex to be:

\begin{equation}
{\Omega}^{ik} = \frac{1}{16 \pi} {{N}^{ikab}}_{,ab}
\end{equation}

where the functions ${N}^{ikab}$ are given by:

\begin{equation}
{N}^{ikab} = \sqrt{-g}\hspace{0.1 cm}[g^{ik} {\eta}^{ab} - g^{ia} {\eta}^{kb} +
g^{ab} {\eta}^{ik} - g^{kb} {\eta}^{ia}]
\end{equation}

and the $\eta^{ab}$ terms represent the components
of a Minkowski metric of signature $-2$.

Once more we use Gauss's theorem to express the total energy as a surface integral:

\begin{equation}
E = \frac{1}{16 \pi} \int \int  {\chi}^{00 \alpha}  {\mu}_{\alpha} dS
\end{equation}

with the superpotentials ${\chi}^{00 \alpha}$ defined as:

\begin{equation}
{\chi}^{00 \alpha} = N^{00k \alpha}_{,k}
\end{equation}

We determine the relevant values of the superpotentials to be:

\begin{eqnarray}
{\chi}^{001} &=& \frac{q^4 + 2 q^2 r^2 + 2 r^4}{r^6}- \frac{2 x(q^2+r^2)^3}{l^2 r^6} \\
{\chi}^{002}&=& \frac{q^4 + 2 q^2 r^2 + 2 r^4}{r^6}- \frac{2 y(q^2+r^2)^3}{l^2 r^6} \\
{\chi}^{003} &=& \frac{q^4 + 2 q^2 r^2 + 2 r^4}{r^6}- \frac{2 z(q^2+r^2)^3}{l^2 r^6}
\end{eqnarray}

Performing the integral in (36) we once again obtain the same energy
distribution found using the other two prescriptions, namely:

 \begin{equation}
E= \frac{q^4}{2 r^3} + \frac{q^2}{r} + \frac{r}{2} -
\frac{(q^2+r^2)^3}{2 l^2 r^3}
\end{equation}

\vspace{1 cm} \noindent
 {\bf 6. Conclusion}

\vspace{0.5 cm} \noindent We have determined that Mart{\'i}nez and
Troncoso's solution representing a static, electrically-charged
black hole with a minimally-coupled self-interacting scalar field
belongs to the Kerr-Schild class.  In applying the Einstein,
Landau-Lifshitz and Papapetrou energy-momentum complexes to the
metric, we have found that each yields an identical local energy
distribution that depends on the electrical charge and cosmological
constant.  Qualitatively this differs from the standard
Reissner-Nordstr{\"o}m solution by the absence of a mass term and
the negative energy value for large radii.

\vspace{1 cm} \noindent
 {\bf Acknowledgements}

\vspace{6 pt} \noindent Thanks to K. S. Virbhadra
for suggesting this problem and for his helpful advice.
\newpage

\vspace{1 cm} \noindent\fontsize{9 pt}{11 pt} \selectfont

\end{document}